\documentclass[letterpaper,11pt,reqno]{article}
\usepackage{paper}
\hypersetup{pdftitle={Minimalist LaTeX Template for Academic Papers}}

\newcommand{\bib}{bibliography.bib}

\usepackage{upgreek,array,soul}

\usepackage{tikz}
\usetikzlibrary{arrows.meta}

\usepackage[labelsep=newline,justification=centering]{caption}

\usepackage{comment}

\begin{document}

% Enter title:
\title{\vspace{-1.25em}Automation Experiments\\and Inequality}

% Enter authors:
\author{%
\begin{tabular}{wc{4cm} c wc{4cm}}
\large{Seth Benzell} & & \large{Kyle R. Myers} \\
\small{Chapman, MIT, \& Stanford} & & \small{Harvard \& NBER} \\ 
& \\
\end{tabular}
%
% Enter affiliations and acknowledgements:
\thanks{Benzell: \href{mailto:benzell@chapman.edu}{benzell@chapman.edu}. Myers: \href{mailto:kmyers@hbs.edu}{kmyers@hbs.edu}. There are no financial relationships or other potential conflicts of interest that apply to any of the authors.}}

% Enter date:
\date{\today} 

% Enter permanent URL (can be commented out):
%\available{}

\begin{titlepage}
\maketitle
% Enter abstract:
An increasingly large number of experiments study the labor productivity effects of automation technologies such as generative algorithms. A popular question in these experiments relates to inequality: does the technology increase output more for high- or low-skill workers? The answer is often used to anticipate the distributional effects of the technology as it continues to improve. In this paper, we formalize the theoretical content of this empirical test, focusing on automation experiments as commonly designed. Worker-level output depends on a task-level production function, and workers are heterogeneous in their task-level skills. Workers perform a task themselves, or they delegate it to the automation technology. The inequality effect of improved automation depends on the interaction of two factors: ($i$) the correlation in task-level skills across workers, and ($ii$) workers' skills relative to the technology's capability. Importantly, the sign of the inequality effect is often non-monotonic --- as technologies improve, inequality may decrease then increase, or vice versa. Finally, we use data and theory to highlight cases when skills are likely to be positively or negatively correlated. The model generally suggests that the diversity of automation technologies will play an important role in the evolution of inequality.
\end{titlepage}

%%%%%%%%%%%%%%%%%%%%%%%%%%%%%%%%%%%%%%%%
%%%%%%%%%% Main text
%%%%%%%%%%%%%%%%%%%%s%%%%%%%%%%%%%%%%%%%%

%%%%%%%%%%%%%%%%%%%%%%%%%%%%%%%%%%%%%%%%
%%%%%%%%%%%%%%%%%%%%%%%%%%%%%%%%%%%%%%%%
\section{Introduction}
%%%%%%%%%%%%%%%%%%%%%%%%%%%%%%%%%%%%%%%%
%%%%%%%%%%%%%%%%%%%%%%%%%%%%%%%%%%%%%%%%

Recent advances in the capabilities of algorithms have led to a surge of  investigations into the economics of automation technologies. An important, open question is the effect of these technologies on earnings inequality. Researchers have approached this question using both macroeconomic and microeconomic methods. In this paper, we are interested in what can and cannot be learned from the microeconomics.\footnote{The macroeconomic work typically asks: after general equilibrium effects play out, will the new distribution of jobs and earnings be less or more unequal? The answer appears often is that technologies that reduce demand for certain types of labor in the short run may increase demand for that labor in the long run, and vice-versa. For overviews covering historical and current waves of automation, see: \cite{katz1999changes,acemoglu2011skills,acemoglu2019automation}.}

Specifically, we focus on the microeconomics of automation experiments: studies that use exogenous variation in the availability of a technology to estimate productivity effects. These experiments are valuable because they yield internally valid estimates (typically, in partial equilibrium) and provide managers and policymakers information about technologies at a relatively high frequency.  In order to estimate inequality effects, experimenters often test whether the technology increases the output of higher- or lower-skilled workers with the same job.\footnote{For examples of automation trials that report inequality results, see: \cite{noy2023experimental,chen2024large,choi2024lawyering,kreitmeir2024heterogeneous,cui2025effects,brynjolfsson2025generative,dell2025knowledge,kanazawa2025ai,kim2025ai,otis2025uneven,roldan2025genai}.} In June of 2025, \emph{The Economist} summarized the inequality effects from the automation experiments to date: ``\emph{Although early studies suggested that lower performers could benefit ... newer studies look at more complex tasks... In these contexts, high performers benefit far more than their lower-performing peers,}'' which implies that the skill \emph{level} of the workforce is an important feature and suggests a future of skill-biased technical change, as implied by the article's title ``\emph{How AI will Divide the Best from the Rest,}'' (\citealt{economist2025ai}). 

But what exactly do we learn if we see an automation experiment lead to more, or less, inequality? What features of the workforce are relevant to the inequality effect? If a technology increases inequality now, will it continue to increase inequality as it becomes more capable?

To answer these sort of questions, we provide a new model of automation experiments where the occupations and workforce are held fixed. The model is in the task-based tradition; output is increasing in task-level inputs. However, unlike many macroeconomic models of task-based production, we explicitly model jobs as production functions at the \emph{individual level}. Individual workers are defined by their endowments of skill in each task, which motivates our focus on a feature of the workforce that is rarely highlighted: the correlation of workers' skills across tasks. This correlation describes, for example, whether the most analytical physicians are also the most empathetic, or whether the best fielder on the baseball team is also the best batter.

We study automation experiments involving either positive or negatively correlated skills among workers. Automation technologies can perform one (or more) of the tasks in production, and a technology's capability may be below or above a given workers' skill. We assume rational expectations, so workers only automate (i.e., delegate to the technology) if the technology's capability is above their skill.\footnote{This assumption is not innocuous, as some automation experiments do find workers to perform worse than placebo when given access to generative algorithms (\citealt{dell2025knowledge,otis2025uneven}), which suggests the presence of biased beliefs.}

The model provides clear predictions about the change in inequality (i.e., the difference in output between worker types) given a marginal increase in an automation technology's capability. Even in simple settings with only two tasks, perfect skill correlations, and single-task automaton, we obtain results that may seem counter-intuitive at first, but become very clear through the lens of the model. Changes in inequality do not depend on the absolute level of workers skill, as the \emph{The Economist} quote above implies or as intuition might suggest. It is the interaction of skill correlation and technological capability that determines the inequality effect. 

For example, consider two samples of workers with identical, two-task Cobb-Douglas production functions and identical, task-levelskill distributions. But in one sample, skills are positively correlated and, in the other, skills are negatively correlated across workers. Here, an increase in the capability of a technology that automates one task can \emph{decrease} inequality in one sample while \emph{increasing} inequality in the other.

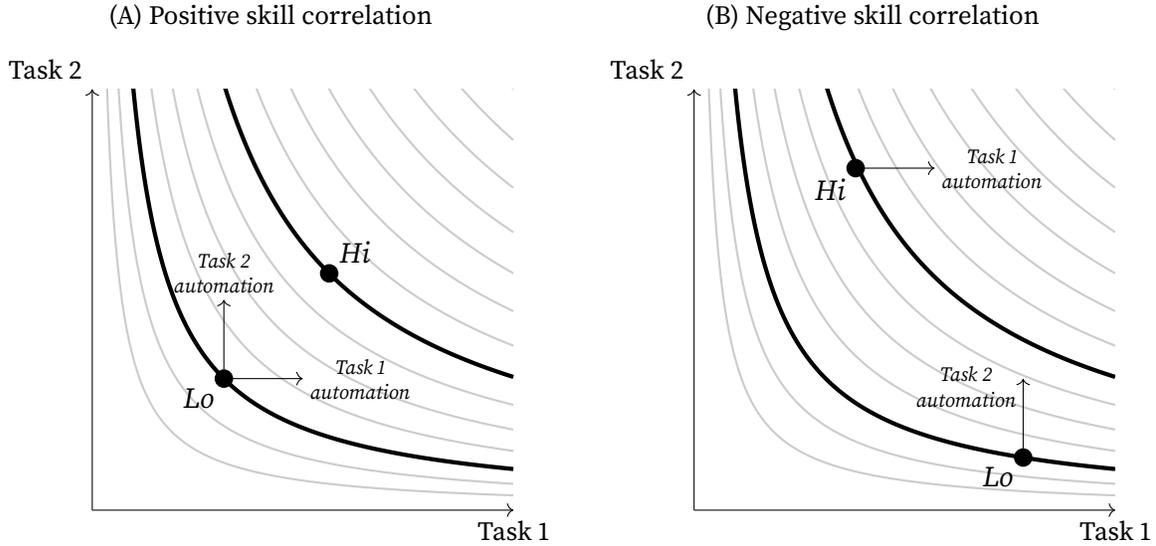
\begin{figure}[h!] 
\caption{Initial Changes in Output after Automation}\label{fig_isoq}
\begin{subfigure}[t]{0.475\textwidth} \caption{Positive skill correlation}\label{fig_isoq_pos}
\begin{tikzpicture}[scale=0.7]
    % Axes
    \draw[->] (0,0) -- (8,0) node[below] {\small Task 1};
    \draw[->] (0,0) -- (0,8) node[above left] {\small Task 2};

    % Keep curves inside a clean 
    \begin{scope}
    \clip (0,0) rectangle (8,8);

    % Isoquants: X2 = (Ybar^2)/X1
    \foreach \Y in {1.5,2,2.5,3,3.5,4,4.5,5,5.5,6,6.5,7,7.5,8,8.5}{
    \draw[thick,black!20!white,domain=0.2:8,samples=300]
        plot (\x,{(\Y*\Y)/\x});
    }

    % Isoquants: X2 = (Ybar^2)/X1
    \foreach \Y in {2.5,4.5}{
    \draw[ultra thick,black,domain=0.2:8,samples=300]
        plot (\x,{(\Y*\Y)/\x});
    }
    
    \fill[black] ({2.5},{(2.5^2)/2.5}) circle (5pt) node[below left,scale=1.0]{$Lo$};
    \draw[->,black] ({2.5},{(2.5^2)/2.5}) -- ++(1.5,0) node[pos=1,right,yshift=0pt,align=center,scale=0.7]{\emph{Task 1}\\\emph{automation}};
    \draw[->,black] ({2.5},{(2.5^2)/2.5}) -- ++(0,1.5) node[pos=1,above,xshift=0pt,align=center,scale=0.7]{\emph{Task 2}\\\emph{automation}};    
    \fill[black] ({4.5},{(4.5^2)/4.5}) circle (5pt) node[above right,scale=1.0]{$Hi$};

    \end{scope}
\end{tikzpicture}
\end{subfigure} \hfill
\begin{subfigure}[t]{0.475\textwidth} \caption{Negative skill correlation}\label{fig_isoq_neg}
\begin{tikzpicture}[scale=0.7]
    % Axes
    \draw[->] (0,0) -- (8,0) node[below] {\small Task 1};
    \draw[->] (0,0) -- (0,8) node[above left] {\small Task 2};

    % Keep curves inside a clean 20x10 window
    \begin{scope}
    \clip (0,0) rectangle (8,8);

    % Isoquants: X2 = (Ybar^2)/X1
    \foreach \Y in {1.5,2,2.5,3,3.5,4,4.5,5,5.5,6,6.5,7,7.5,8,8.5}{
    \draw[thick,black!20!white,domain=0.2:8,samples=300]
        plot (\x,{(\Y*\Y)/\x});
    }

    % Isoquants: X2 = (Ybar^2)/X1
    \foreach \Y in {2.5,4.5}{
    \draw[ultra thick,black,domain=0.2:8,samples=300]
        plot (\x,{(\Y*\Y)/\x});
    }
    
    % \fill[black] ({1},{(2.5^2)/1}) circle (5pt) node[below left,scale=1.0]{$L$};
    % \draw[->,black] ({1},{(2.5^2)/1}) -- ++(1.4,0) node[pos=1,below,yshift=-3pt,align=center,scale=0.7]{\emph{Task 1}\\\emph{automation}};
    \fill[black] ({(2.5^2)/1},{1}) circle (5pt) node[below left,scale=1.0]{$Lo$};
    \draw[->,black] ({(2.5^2)/1},{1}) -- ++(0,1.5) node[pos=1,left,yshift=-3pt,align=center,scale=0.7]{\emph{Task 2}\\\emph{automation}};
    
    % \fill[black] ({6.5},{(4.5^2)/6.6}) circle (5pt) node[below left,scale=1.0]{$H$};
    % \draw[->,black] ({6.5},{(4.5^2)/6.6}) -- ++(0,1.5) node[pos=1,above,xshift=0pt,align=center,scale=0.7]{\emph{Task 2}\\\emph{automation}};
    \fill[black] ({(4.5^2)/6.6},{6.5}) circle (5pt) node[below left,scale=1.0]{$Hi$};
    \draw[->,black] ({(4.5^2)/6.6},{6.5}) -- ++(1.5,0) node[pos=1,right,xshift=0pt,align=center,scale=0.7]{\emph{Task 1}\\\emph{automation}};
    \end{scope}
\end{tikzpicture}
\end{subfigure}  
\note{\emph{Note:} Plots isoquants of production possibilities, highlighting the isoquant for a worker with higher output ($Hi$) and a worker with lower output ($Lo$) in two scenarios, where task-level skills are positively correlated (Panel A) or negatively correlated (Panel B). Workers pre-automation endowments are shown by the black circles. If a task-specific automation technology is made available and the technology's capability exceeds the workers skill endowment, they will use the technology, which will shift them to a new production frontier as shown by the arrows. For illustrative purposes, the first worker to benefit from automation of a given task is highlighted.}
\end{figure}

To illustrate the intuition of this result, consider Figure \ref{fig_isoq}, which reflects this example. In both cases, the production function and the initial level of inequality (per the difference in output between high-type $H$ and low-type $L$) is the same; however, in Panel A, skills are positively correlated while in Panel B skills are negatively correlated. As, for instance, Task 1 automation technology improves, the first user will be the low type in the positive correlated case, but it will be the high type in the negative correlation case. Thus, at least initially, the low type will benefit first and inequality will decrease in the Panel A case, but the high type will benefit first and inequality will increase in the Panel B case. 

Another interesting result is that the inequality effect need not be monotonic in the automation technology's capability. Continuing with the example in Figure 1 Panel A, let's again focus on Task 1 automation. As just noted, the low type will be the initial users, which decreases inequality initially. However, once the technology's capability surpasses the high type's skill and the high type begin usage, output differences will be governed by differences in Task 2 skills. Now, since the high type is more skilled at Task 2 (and tasks are complementary), inequality will \emph{increase} as the Task 1 automation technology improves.

Allowing for multi-task automation yields more interesting patterns. The key result we find is that the shortest path to equality is when automation improvements are balanced across tasks. This motivates the potential importance of a diversity of technological advances when it comes to the pursuit of equality. But importantly, ``equality'' in our model is possible only because workers are homogeneous with respect to the non-automatable skills and we allow for the possibility that technologies' capabilities surpass all workers. Thus, automation may ultimately still lead be inequality in practice.

While several dynamics are possible, ours is not an ``anything goes'' model. When information is available about workers' task-level skills and the relative capability of the automation technology, our model can be use for predicting or extrapolating inequality effects. Thus, it should prove useful for researchers, organizations, managers and policymakers as they interpret and design future automation experiments.

While we focus on partial equilibrium and microeconomics, our model suggests some new insights into the macroeconomics of automation. In particular, the model's focus on skill correlations emphasizes the importance of understanding whether the ``experiment'' is occurring at the job-, organization-, sectoral-, or societal-level. Existing theory and data provide some guidance on when and where we should expect skill correlations to be positive or negative. We discuss this in the latter portion of the paper and report some country-level skill correlations (based on SAT scores) from the US National Longitudinal Survey of Youth (\cite{bls2024nlsy97}). We also consider a simplified version of \citeauthor{kremer1993ring}'s (\citeyear{kremer1993ring}) O-ring model of production to show how skills can be negatively correlated within firms even if they are positively correlated in the population.

After a brief literature review below, the paper is organized as follows: Section \ref{sec_model_setup} lays out the model in detail; Section \ref{sec_model_results} derives results from the model; Section \ref{sec_empiric} connects our theoretical results to evidence on skill correlations across the economy and how the level of aggregation (e.g., firms versus populations) has an effect on these correlations; and Section \ref{sec_discuss} concludes with a discussion of our model's implications, limitations, and possible extensions.

\subsection*{Related Literature}\label{lit}

Our research connects to the literatures on the productivity impact of emerging technologies, most notably, generative algorithms, theories that model jobs as multi-task production functions, and the broader economics of automation. 

\paragraph{Generative Algorithms and Inequality} Table \ref{tab_litreview} summarizes the range of recent automation experiments that have estimated inequality effects due to generative algorithms. As the table shows, the evidence is decidedly mixed. Our theoretical framework helps rationalize these seemingly inconsistent findings. Given the rapid pace of algorithmic progress, experimental results at one point in time may not predict effects as technologies continue to advance. Moving forward, experiments that measure not only productivity effects but also the task-level skill correlations among workers and the technology's capability relative to worker skills will provide more robust guidance for anticipating distributional consequences.

\begin{table}[h!] \centering
\caption{Inequality Results in Recent Automation Experiments}\label{tab_litreview}
{\renewcommand{\arraystretch}{1.25} \footnotesize %
\begin{tabular}{r c c c}
\hline\hline
& \multirow{2}*{\emph{job}} & \multirow{2}*{\emph{methodology}} & \multirow{2}*{$\substack{{ \text{\footnotesize \emph{inequality}} }\\{ \text{\footnotesize \emph{effect}} }}$} \\
& \\
\hline
\\ [-1em]
\cite{chen2024large}& Ad copywriting & RCT & $\downarrow$ \\
\cite{choi2024lawyering} & Legal work & RCT & $\downarrow$ \\
\cite{cui2025effects} & Coding & RCT & $\downarrow$ \\
\cite{dell2025knowledge} & Consulting & RCT & $\downarrow$ \\
\cite{noy2023experimental} & Professional writing & RCT & $\downarrow$ \\
\cite{brynjolfsson2025generative} & Customer support & Natural Experiment & $\downarrow$ \\
\cite{kanazawa2025ai} & Taxi driving & Natural Experiment & $\downarrow$ \\
[1em]
\cite{kim2025ai} & Investment analysis & RCT & $\uparrow$ \\
\cite{otis2025uneven} & Entrepreneurship & RCT & $\uparrow$ \\
\cite{roldan2025genai} & Debating & RCT & $\uparrow$ \\
[0.5em]
\hline\hline
\end{tabular}}
\note{\emph{Notes:} Summarizes recent automation experiments involving generative algorithms. Studies are sorted based on whether they find automation to decrease ($\downarrow$) or increase ($\uparrow$) inequality, whether they use a formal randomized controlled trial (RCT) or a natural experiment, and then alphabetically.}
\end{table}

\paragraph{Jobs as Production Functions} We are not the first to treat jobs as production functions. One close example is \citeauthor{autor2013putting} (\citeyear{autor2013putting}; see their Eq. 2), which models job-level output as an exponential function of workers' task-level skills and job-task-specific elasticities. There, the focus is not on automation, but rather on predicting job-level sorting. \cite{deming2017growing} considers the complementarity of workers' math and social skills, but the correlation in these skills across workers plays no role. \cite{dessein2006adaptive} also consider jobs as bundles of tasks, although the focal question is how organizations endogenously decide which tasks are bundled together. Our model's highlighting of task-level correlation in skills should prove useful in future work using worker-level production functions, specifically, and the division of labor more generally (e.g., \citealt{becker1992division,autor2003skill}). We review empirical studies of skill correlations later in Section \ref{subsec_empiric_litreview}.

\paragraph{Automation in General} Related work in the broader economics of automation includes \cite{athey2020allocation}, who develop a framework for optimal delegation between humans and algorithms, and \cite{xu2025generative}, who study how generative algorithms affects organizational structure and the allocation of tasks within firms. \cite{trammell2025workflows} focuses on the automation of tasks when there is learning-by-doing in a sequence of tasks that constitute the workflow for a job. Regarding the macroeconomics of automation and inequality, \cite{katz1999changes}, \cite{acemoglu2011skills}, and \cite{acemoglu2019automation} provide comprehensive overviews of how technological change affects the wage structure through general equilibrium channels. More recently, \cite{autor2025expertise} examines how generative algorithms affects the returns to expertise, while \cite{garicano2025training} studies implications for training and apprenticeship. Our partial equilibrium approach abstracts from many of these sort of general equilibrium effects in order to isolate the direct productivity channel, which yields predictions about which workers benefit from automation in the short run before labor markets adjust. Extensions of leading general equilibrium models of automation (e.g., \citealt{zeira1998workers,acemoglu2018race}) to incorporate task-level correlations may prove fruitful.

%%%%%%%%%%%%%%%%%%%%%%%%%%%%%%%%%%%%%%%%
%%%%%%%%%%%%%%%%%%%%%%%%%%%%%%%%%%%%%%%%
\section{Model Preliminaries}\label{sec_model_setup}
%%%%%%%%%%%%%%%%%%%%%%%%%%%%%%%%%%%%%%%%
%%%%%%%%%%%%%%%%%%%%%%%%%%%%%%%%%%%%%%%%
Here, we outline our model of jobs as production functions and the automation of tasks. The tasks of the production functions and the skills of workers are exogenous and fixed. This short-run, partial-equilibrium view reflects the format of most automation experiments given their pursuit of internal validity.

%%%%%%%%%%%%%%%%%%%%%%%%%%%%%%%%%%%%%%%%
\subsection{Jobs as Production Functions}
%%%%%%%%%%%%%%%%%%%%%%%%%%%%%%%%%%%%%%%%
There is a single job where each worker $i$ produces a non-negative quantity of some output $Y$. The job involves a set of tasks $t=\{1,2,...,T\}$, and workers inelastically supply a unit of effort towards each task.\footnote{In other words, workers are endowed with an amount of labor in each task. An equivalent framing is that ``time-on-task'' is fixed. Allowing for endogenous time-on-task, may change the absolute magnitudes of output levels. But with rational expectations, workers will still allocate their time in a way that is positively correlated with their relative skill levels. Thus, we expect the levels of inequality, and the exact points at which the sign of the inequality effects change, to depend on endogeneity of time-on-task. But we do not expect the non-monotonicity results we obtain below to become monotonic.}  Workers are differentiated by their task-specific skills $L_{it} >0$. Since labor is inelastic, we can describe each worker's exogenous (job-specific) skill endowment across tasks as a vector $\textbf{\textup{L}}_i=(L_{i1}, L_{i2}, ..., L_{iT})$.

Output is produced according to a general CES production function:
\begin{equation}
    Y_i = \Bigl(\sum_t \alpha_t L_{it}^\rho  \Bigr)^{1/\rho} \qquad \text{where} \qquad \rho=\frac{\sigma-1}{\sigma}\leq 1\,,\;
\sigma \geq 0 \,,\; \sum_t \alpha_t=1 \;,
\end{equation}
where $\sigma$ is the job-specific elasticity of substitution, and $\alpha_t$ are the task-specific input shares. For simplicity, we assume the job has two tasks ($T=2$) with equal input shares ($\alpha_t=1/2$). This simplifies the production function to:
\begin{equation}
    Y_i = \Bigl(\frac{1}{2} \bigl( L_{i1}^\rho + L_{i2}^\rho \bigr) \Bigr)^{1/\rho} \;.
\end{equation}

For some results, we focus on the edge case of $\rho \rightarrow 0$, which yields the familiar Cobb-Douglas production function: 
\begin{equation}
    Y_i = ( L_{i1} \times L_{i2})^{1/2} \;.
\end{equation}

%%%%%%%%%%%%%%%%%%%%%%%%%%%%%%%%%%%%%%%%
\subsection{Automation Experiments}
%%%%%%%%%%%%%%%%%%%%%%%%%%%%%%%%%%%%%%%%
We model automation experiments as the exogenous offering of a technology that can automatically perform a task. We denote the capability of the automating technology for task $t$ as $A_t \geq0$. As an input, the automating technology may be worse, equal to, or better than labor (i.e., $A_t \lesseqgtr L_{it}$). We assume that workers have rational expectations about the technology's capability; thus, $i$ will rationally choose to use it if $A_t > L_{it}$ and not use it otherwise.

The production function with single-task automation becomes:
\begin{equation}
    Y_i = \Bigl(\frac{1}{2} \bigl( \max(L_{it}, A_t)^\rho + L_{it^-}^\rho \bigr) \Bigr)^{1/\rho} \;.
\end{equation}
where $t$ is the task being automated and $t^-$ is the other task. When we consider cases where both tasks are being automated, we will use the max function for both tasks.

%%%%%%%%%%%%%%%%%%%%%%%%%%%%%%%%%%%%%%%%
\subsection{Skill Correlations}
%%%%%%%%%%%%%%%%%%%%%%%%%%%%%%%%%%%%%%%%
We consider the two limit cases for the skill correlation structure, both in which there is ex-ante (pre-experiment) differences in output levels. There are two groups of workers: those with initially high output ($i=H$) and those with initially low output ($i=L$).\footnote{This binary categorization is simple, and it reflects the common empirical exercise of proxying for workers' skills using their pre-experiment output levels to divide them into two groups (e.g., \citealt{noy2023experimental,chen2024large,choi2024lawyering,cui2025effects,brynjolfsson2025generative,dell2025knowledge,kanazawa2025ai,kim2025ai,otis2025uneven,roldan2025genai}).} Table \ref{tab_abilitycorr} summarizes our negative and positive correlation cases, noting that we use the variables $C$ and $B$ to describe the workers comparative and/or absolute advantages, where $1<C<B$.\footnote{Note this analytical setup differs slightly from the data-generating process used to produce Figure \ref{fig_isoq}. There, the focus was on ease of visual understand, and here the focus is on analytical tractability.}

\begin{table}[h]\centering
\caption{Skill Correlation Scenarios --- Worker ($i$) and Job Task ($t$)}\label{tab_abilitycorr}
{\renewcommand{\arraystretch}{1.25} \footnotesize%
\begin{tabular}{c c c c c c }
    \hline\hline
    & \multicolumn{2}{c}{\emph{positive corr.}} & & \multicolumn{2}{c}{\emph{negative corr.}} \\ \cline{2-3} \cline{5-6}
    & $t=1$ & $t=2$ & & $t=1$ & $t=2$ \\
    \hline
    \\ [-1em]
    $i=H$ & $B$ & $BC$ & & $B$ & 1 \\
    $i=L$ & 1 & $C$ & & 1 & $C$ \\
    [0.5em]
     \hline\hline
\end{tabular}}
\note{\emph{Notes}: Reports the values of workers' task-specific skills for the positive and negative correlation cases; $1<C<B$.}
\end{table}

%%%%%%%%%%%%%%%%%%%%%%%%%%%%%%%%%%%%%%%%
\subsection{Inequality Measures and Effects}
%%%%%%%%%%%%%%%%%%%%%%%%%%%%%%%%%%%%%%%%
There are many ways to measure inequality. In practice, one of the most common ways worker-level output inequality is evaluated is simply the difference in output levels between workers. The change in inequality is estimated by testing whether workers with high initial output levels (pre-automation) benefit more or less than workers with low initial output levels. If those initially high-output workers benefit more, inequality is said to have increased, and vice versa.

As one way of formalizing this, first define $\Delta$ as:
\begin{equation*}
    \Delta \equiv Y_H(A_t) - Y_L(A_t) \;,
\end{equation*}
which provides a directional measure of inequality --- below, we'll also investigate the absolute value of $\Delta$ as a measure of inequality that is agnostic as to the source of the differences. This leads us to our main (partial equilibrium) effect of interest:
\begin{equation*}
    \partial \Delta / \partial A_t\;,
\end{equation*}
which describes how the output gap between workers of groups $H$ and $L$ changes as automation technologies are introduced and/or improved.

As we discussed in the introduction, many automation experiments motivate a focus on $\partial \Delta / \partial A_t$ with concerns about long-run general equilibrium effects (or, perhaps a latent preference over short-run inequality). A straightforward critique of this motivation is that, in the pursuit of internal validity, many of these experiments hold the worker population and nature of the job fixed --- this rules out general equilibrium adjustments (e.g., workers changing jobs, or employers changing the task composition of jobs). 

One intuitive rebuttal to these critiques is to make the assumption that the magnitude or the of the inequality effect ($\partial \Delta / \partial A_t$), should be persistent and informative of the potential for the technology to affect inequality. Formally, this could stated as an assumption that $\partial \Delta / \partial A_t$ is constant, or at least the same sign, as $A_t$ increases. We investigate this possibility by solving for the second derivative, $\partial^2 \Delta / \partial A^2$.

To further explore heterogeneity, we also solve for the ways in which the inequality effect depends on: ($i$) how large the absolute advantage of the high-type workers is, $\partial^2 \Delta / \partial A \partial (B/C)$; and ($ii$) how substitutable the two tasks are in the production function, $\partial^2 \Delta / \partial A \partial \rho$. Furthermore, we'll investigate the case where both tasks may be automated and illustrate how the effect of inequality jointly depends on the capabilities of the two technologies, $\partial^2 \Delta / \partial A_1\partial A_2$.

Note that $\Delta$ does not measure \emph{absolute} inequality. For instance, it could be the case that automation differentially leads the low type workers to produce so much more output that the absolute difference in output between workers grows beyond the initial level. Thus, we also focus our attention on the absolute value of $\Delta$:
\begin{equation*}
    |\Delta| \equiv |Y_H(A_t) - Y_L(A_t)| \;.
\end{equation*}

Across the literature, other common metrics of inequality includes statistics such as the coefficient of variation or the gini coefficient. These metrics typically involve a scaling of absolute differences by some measure of aggregate levels.\footnote{For example, in our setting, the coefficient of variation is simply: $|Y_H-Y_L|/(Y_H+Y_L)$; and the gini coefficient is simply the coefficient of variation divided by 2.} This scaling is useful for making comparisons across contexts, since it converts absolute values into relative magnitudes. However, since we will always be considering the same context in theory, we will rely on our simpler measures.

%%%%%%%%%%%%%%%%%%%%%%%%%%%%%%%%%%%%%%%%
%%%%%%%%%%%%%%%%%%%%%%%%%%%%%%%%%%%%%%%%
\section{Model Results}\label{sec_model_results}
%%%%%%%%%%%%%%%%%%%%%%%%%%%%%%%%%%%%%%%%
%%%%%%%%%%%%%%%%%%%%%%%%%%%%%%%%%%%%%%%%

%%%%%%%%%%%%%%%%%%%%%%%%%%%%%%%%%%%%%%%%
\subsection{Single-task Automation}
%%%%%%%%%%%%%%%%%%%%%%%%%%%%%%%%%%%%%%%%

\begin{figure}[h!] \centering
\caption{Inequality and Single-task Automation --- Cobb-Douglas Production}\label{fig_deltas}
\begin{subfigure}[t]{0.475\textwidth} \caption{Positive skill correlation}\label{fig_deltas_pos}
\includegraphics[width=\textwidth, trim=0mm 20mm 0mm 0mm, clip]{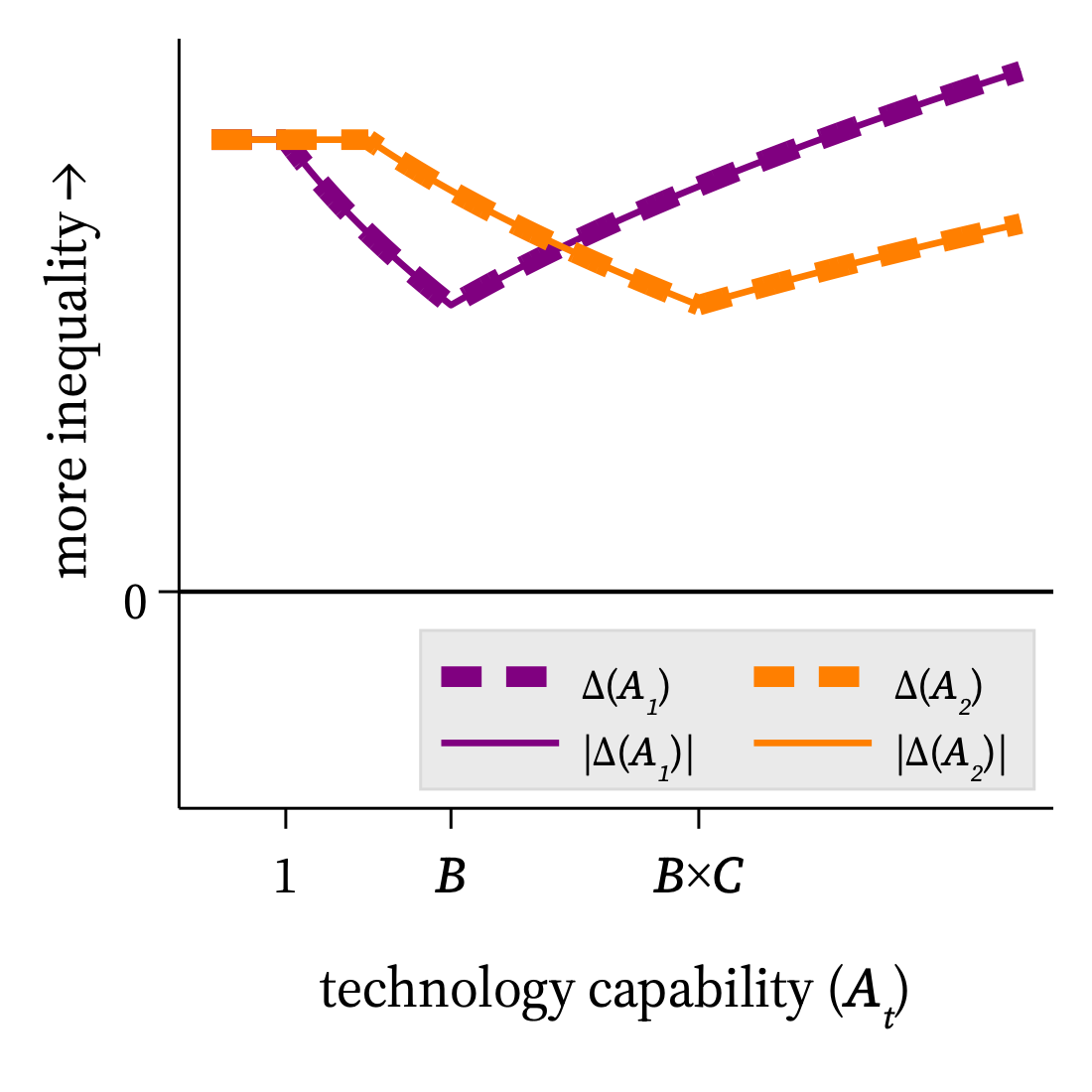}
\end{subfigure} \hfill
\begin{subfigure}[t]{0.475\textwidth} \caption{Negative skill correlation}\label{fig_deltas_neg}
\includegraphics[width=\textwidth, trim=0mm 20mm 0mm 0mm, clip]{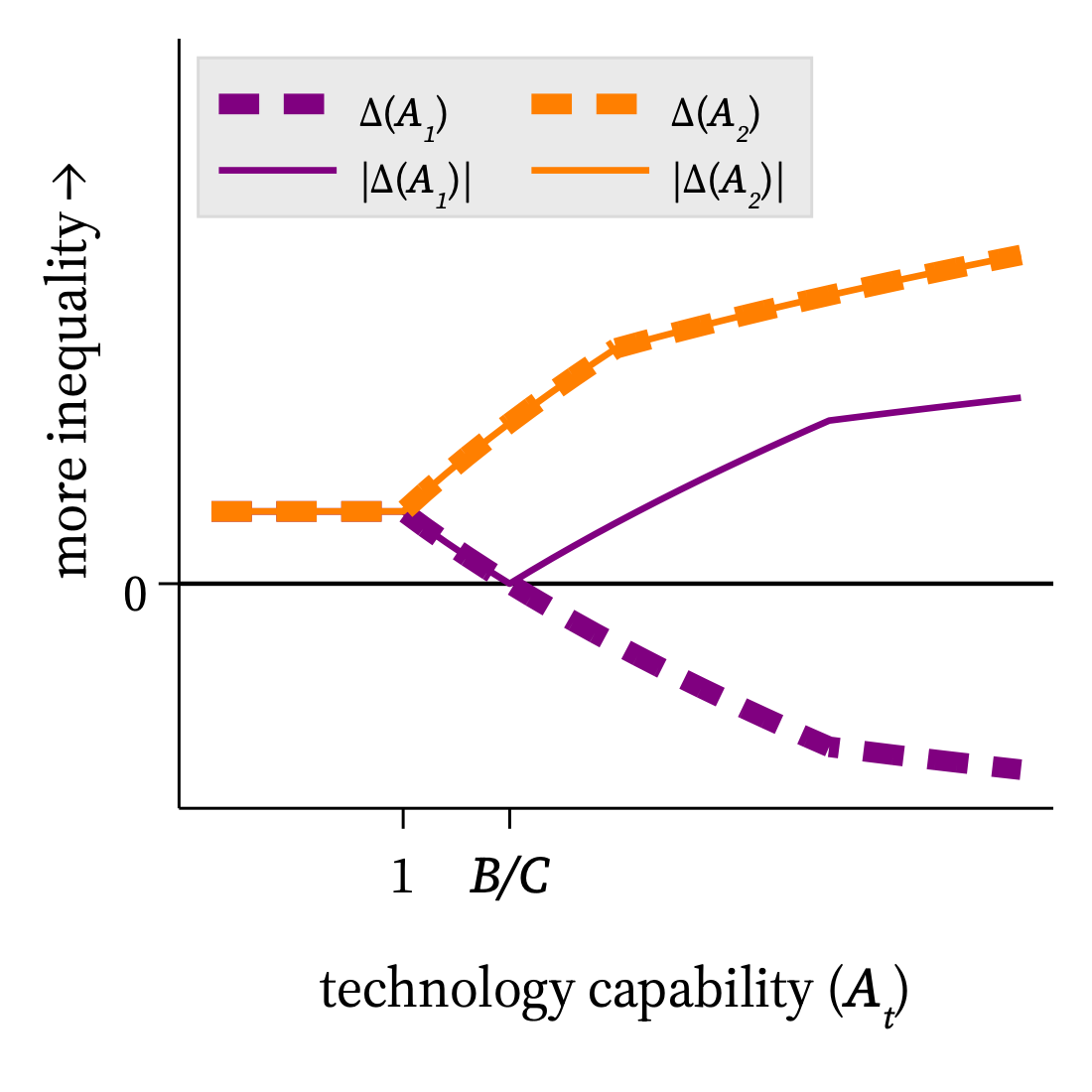}
\end{subfigure}
\note{\emph{Note:} Plots the level of inequality across the two groups of workers assuming Cobb-Douglas production per relative inequality ($\Delta$; solid line) or absolute inequality ($|\Delta|$; dashed line) depending on skill correlations, the task being automated, and the capability of the technology.}
\end{figure}

We begin with a focus on the simple case of Cobb-Douglas production and single-task automation. Figure \ref{fig_deltas} summarizes the results visually; it illustrates relative and absolute inequality (per line pattern) as a function of the correlation in workers' skills (per Panel), which task is automated (per line color), and the capability of the automation technology (per the $x$-axis).

When workers' skills are positively correlated, relative and absolute inequality measures always align (see Figure \ref{fig_deltas_pos}). Furthermore, regardless of whether task 1 or 2 is automated, there will be a non-monotonic relationship between inequality and the automation technology's capability; first inequality declines, then it increases. Initially, the automation helps the low type ``catch-up'' to the high-output type. This is because the technology's capability is greater than the low type (so, the low type uses the tool), but lower than the high type (so, the high type relies on their own skill). But eventually, the technology is better than both types, and the high-types absolute advantage in the non-automated task becomes the key determinant of output differences and inequality grows due to the complementarity in the tasks.

In the case of negatively correlated skills, automating task 1 reduces \emph{relative} inequality ($\Delta$) for all technology capability levels, while automating task 2 increases relative inequality for all technology capabilities (see Figure \ref{fig_deltas_neg}). Absolute inequality ($|\Delta|$) always increases with improvements in task 2 automation; however, this monotonicity does not occur for absolute inequality as task 1 automation improves. Initially, absolute declines, but when the technology's task 1 capability ($A_1$) equals $B/C$ absolute inequality begins to grow again. At this point, the initially low-output worker types benefit so much from the automation of task 1 (the task they have both an absolute and comparative disadvantage in) that their output exceeds that of the initially high-output worker. This divergence of relative and absolute inequality highlights the importance of tracking workers over time in this particular scenario. For example, cross-sectional analyses of automation experiments may reveal no change in inequality that mask a re-ordering of workers.

Clearly, the specific points at which these non-monotonic switches in inequality occur will be context-specific. But overall, the simple Cobb–Douglas case clearly illustrates how improvements in automation may increase or decrease output inequality depending on the workers' skill correlations and the capability of the technology relative to each workers' skills. 

\begin{table}[h!] \centering
\caption{Absolute Output Inequality and Automation}\label{tab_cesabs}
{\renewcommand{\arraystretch}{1.25} \footnotesize %
\begin{tabular}{cc c c c c c c c c } 
    \hline\hline
    \multirow{2}*{\emph{technology}} & \multirow{2}*{\shortstack[c]{\emph{tech.}\\ \emph{capability}}} & & \multirow{2}*{$sign\,\frac{\partial |\Delta|}{\partial A}$} & & \multirow{2}*{$sign\,\frac{\partial^2 |\Delta|}{\partial A^2}$} & & \multirow{2}*{$sign\,\frac{\partial^2 |\Delta|}{\partial A \partial B/C}$} & & \multirow{2}*{$sign\,\frac{\partial^2 |\Delta|}{\partial A \partial \rho}$} \\ 
    & \\
    \hline
    \\ [-1em]
    \multicolumn{10}{l}{\emph{\underline{Panel (A): Negative skill correlation}}} \\
    $A_1$ & $1 < A_1 < A^*$  & & $(-)$ & & $(+)$ & & 0 & & $(+)$ \\
    $A_1$ & $A^* < A_1 $  & & $(+)$ & & $(-)$ & & 0 & & $(-)$ \\
    [1em]
    $A_2$ & $1 < A_2$  & & $(+)$ & & $(-)$ & & 0 & & $(-)$ \\
    [1.5em]
    \multicolumn{10}{l}{\emph{\underline{Panel (B): Positive skill correlation}}} \\
    $A_1$ & $1 < A_1 < B$  & & $(-)$ & & $(+)$ & & 0 & & $^\text{\color{white}a}(\pm)^\text{a}$ \\
    $A_1$ & $B < A_1 $  & & $(+)$ & & $(-)$ & & $(+)$ & & $(-)$ \\
    [1em]
    $A_2$ & $C < A_2 < BC$  & & $(-)$ & & $(+)$ & & 0 & & $(-)$ \\
    $A_2$ & $BC < A_2$  & & $(+)$ & & $(-)$ & & $(+)$ & & $(-)$ \\
    [0.5em]
    \hline\hline
\end{tabular}}
\note{\emph{Notes:} Reports the sign of the first and second derivatives of the absolute inequality effect ($\partial |\Delta|$) for the general CES production function with automation depending on whether workers' skills are negatively correlated (Panel a) or positively correlated (Panel b). $A^{*}=(B^{\rho}+1-C^{\rho})^{1/\rho}$. $(\pm)^\text{a}$: sign is positive if $A<C$ and negative if $C<A<B$.}
\end{table}

Table \ref{tab_cesabs} generalizes these results to the CES production function. The table reports the signs of first and second derivatives of absolute inequality with respect to automation capability, providing a more complete picture of when and why automation's inequality effects may reverse.

In all cases, the sign of this second derivative with respect to the technology's capability ($\partial^2 |\Delta| / \partial A^2$) is opposite to the sign of the first derivative, which stems from the concavity of the production function. This matters for practice: a firm piloting automation today may observe inequality changing, and even before the dramatic non-monotonicity occurs, the rate of change in inequality will decline. 

The cross-derivatives with respect to the high-type's skill advantage ($\partial^2 |\Delta| / \partial A \partial B/C$) reveal a few instances where initial skill gaps can amplify automation's inequality effects. In the negative skill correlation case, the skill gap ($B/C$) plays no role in the inequality effect of automation. However, this derivative is positive in the positive correlation case once technology's capabilities exceeds both workers' skills --- a larger pre-existing skill gap means automation will increase inequality more sharply. For organizations considering automation, this implies that the same technology will have different distributional consequences depending on how heterogeneous the workforce is to begin with. A narrow skill distribution may mute inequality changes; a wide distribution may amplify them.

The cross-derivatives with respect to task complementarity ($\partial^2 |\Delta| / \partial A \partial \rho$) indicate whether automation's effect on inequality is amplified or not in jobs with more substitutable tasks ($\rho \rightarrow 1$). In the case of negative skill correlation, task substitutability behaves just as the second derivative; more substitutability mutes the inequality effect. With positively correlated skills, the cross derivative is negative for most technological capabilities, which amplifies inequality effects at low capability levels but depresses them at higher capability levels.

Practically, these comparative statics tell us that predicting automation's inequality effects requires knowing not just current worker skills and technology capabilities, but also the job's production structure and the trajectory of technological improvement.  Interestingly, these exercises of improving technology's capabilities for a single task at a time all predict that inequality will, in partial equilibrium, eventually increase in the long-run (i.e., $A_t \rightarrow \infty$).

%%%%%%%%%%%%%%%%%%%%%%%%%%%%%%%%%%%%%%%%
\subsection{Multi-task Automation}
%%%%%%%%%%%%%%%%%%%%%%%%%%%%%%%%%%%%%%%%

\begin{figure}[h!] \centering
\caption{Absolute Inequality and Multi-task Automation --- Cobb-Douglas Production}\label{fig_deltas_bothA}
\begin{subfigure}[t]{0.475\textwidth} \caption{Positive skill correlation}\label{fig_deltas_bothA_pos}
\includegraphics[width=\textwidth, trim=0mm 20mm 0mm 0mm, clip]{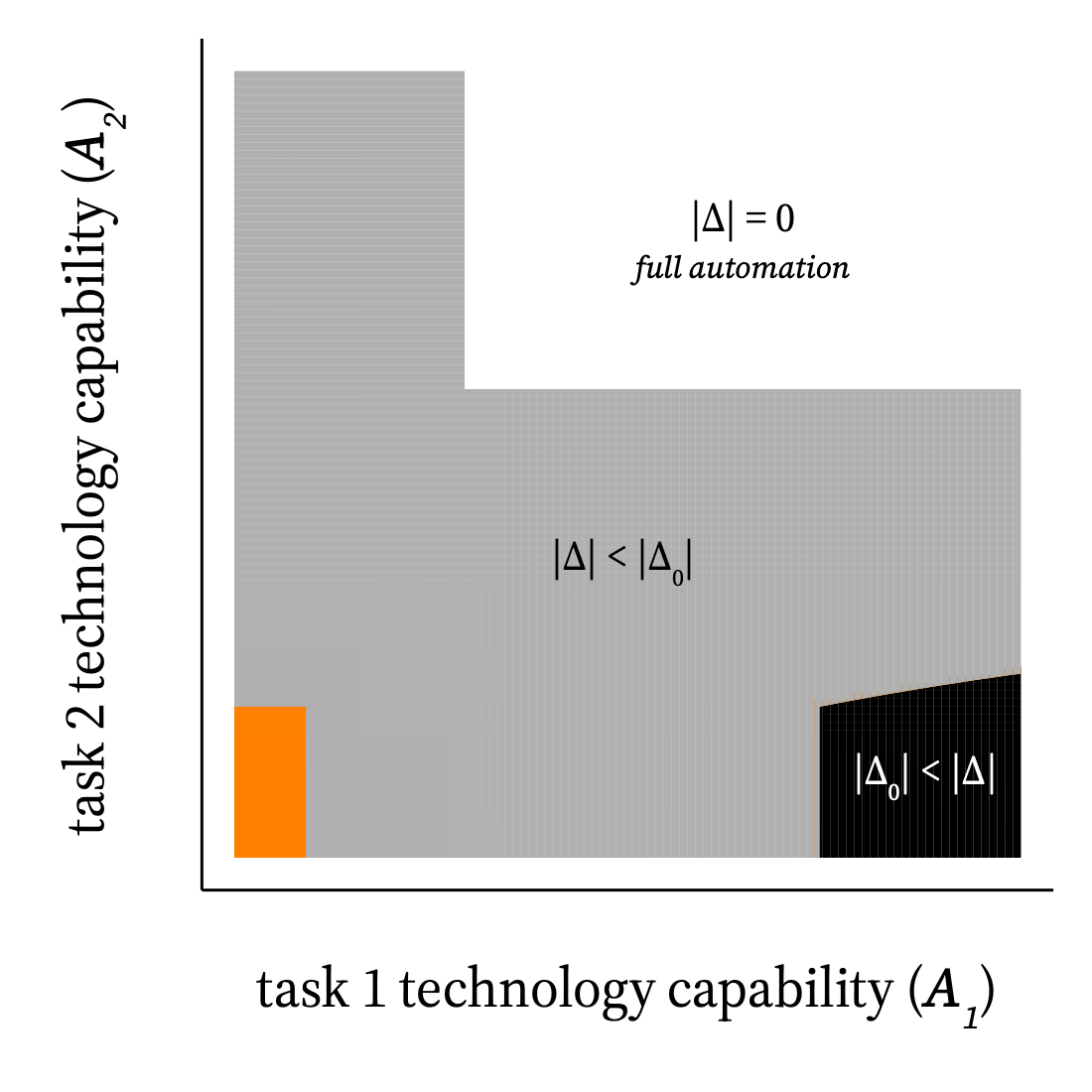}
\end{subfigure} \hfill
\begin{subfigure}[t]{0.475\textwidth} \caption{Negative skill correlation}\label{fig_deltas_bothA_neg}
\includegraphics[width=\textwidth, trim=0mm 20mm 0mm 0mm, clip]{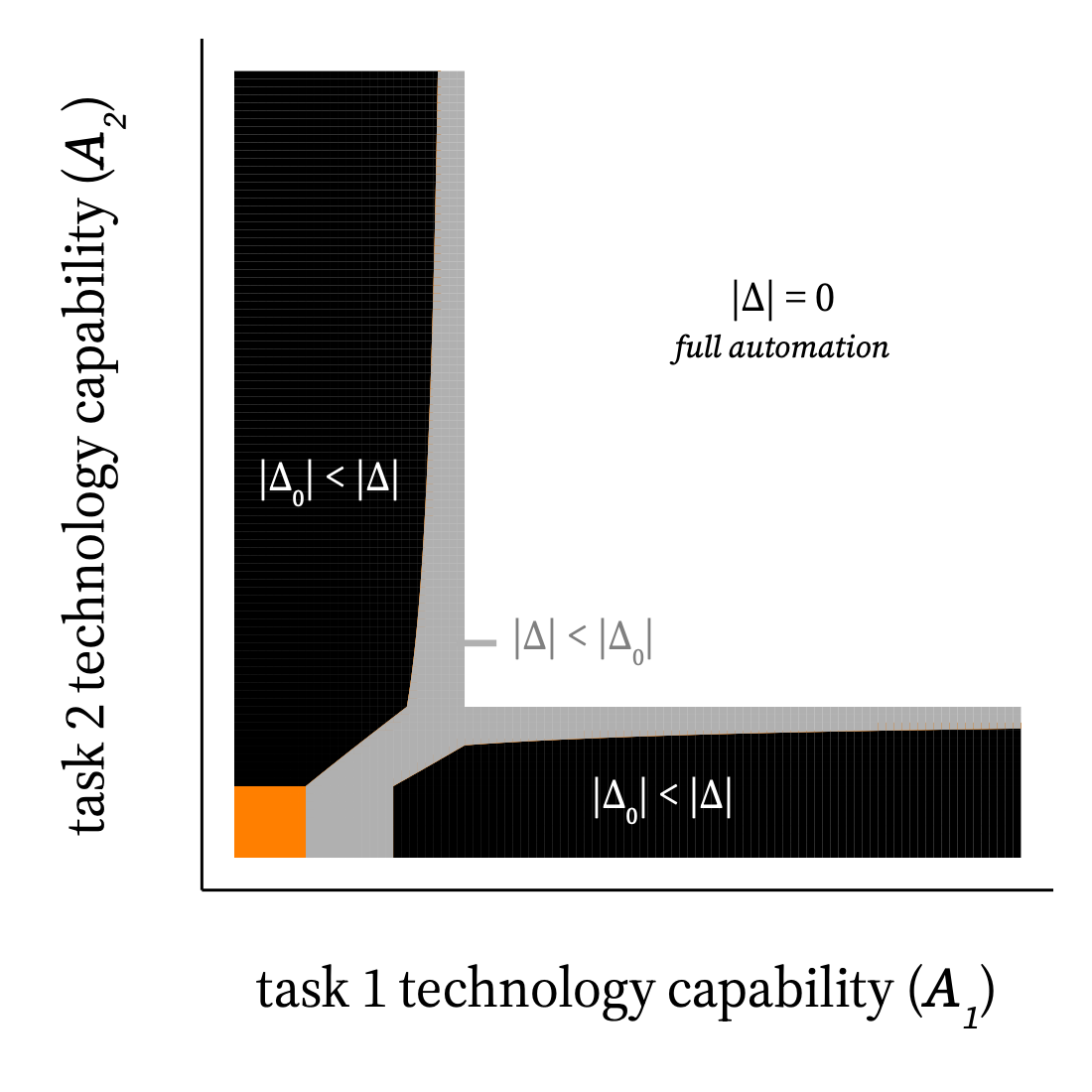}
\end{subfigure}
\note{\emph{Note:} Plots the level of absolute inequality ($|\Delta|$) compared to the initial, pre-automation level of absolute inequality (defined here as: $|\Delta_0|$) depending on skill correlations and the capability of both automation technologies. The results here assume Cobb-Douglas production, and the orange areas indicate regions where neither technology has been adopted (i.e., $|\Delta|=|\Delta_0|$).}
\end{figure}

Returning to the Cobb-Douglas formulation for simplicity, Figure \ref{fig_deltas_bothA} extends the analysis to consider automation of both tasks, revealing how the \emph{composition} of automation technologies shapes inequality outcomes. Each panel plots absolute inequality compared to the pre-automation period as a function of both technologies' capabilities. The contrast between panels highlights how skill correlations determine not just whether inequality rises or falls, but which combinations of technologies prove most equalizing.

In both cases, there is convergence to zero inequality when automation is sufficiently advanced; however, we note again the fact that our model assumes workers are homogeneous with respect to non-automatable tasks (since there are none in our model). This is an important abstraction. Until the economy achieves full automation, there will be non-automated tasks, so the ``equality'' we arrive at should be understood as representing the amount of inequality determined by all non-automated tasks.

With that caveat in mind, Figure \ref{fig_deltas_bothA} reveals the asymmetric paths to equality. When automation advances differentially across tasks, inequality remains or grows --- the technology portfolio is unbalanced relative to workers' skill distributions. This illustrates how the balance of technological capabilities is relevant to inequality. As more tasks are automated, the dimensionality along which workers differ shrinks, and inequality increasingly reflects skill differences in the remaining non-automated tasks. 

With positive correlation (see Figure \ref{fig_deltas_bothA_pos}), much of the capability space involves decreases in inequality compared to pre-automation periods. Effectively, technologies are always helping the low type catch up. Only if the task 2 technology has very low capability and task 1 technology has very high capability can inequality exceed the pre-automation period.

With negative correlation (see Figure \ref{fig_deltas_bothA_neg}), there is a much greater opportunity for increases in inequality. The intuition follows from the single-task results: when skills are negatively correlated, different workers are disadvantaged in different tasks, so equalizing outcomes requires helping both types of workers. A technology suite that is very capable in task 1 helps low-types, leaving low-types' comparative advantage in task 2 intact, which remains a (potential) source of inequality.

%%%%%%%%%%%%%%%%%%%%%%%%%%%%%%%%%%%%%%%%
%%%%%%%%%%%%%%%%%%%%%%%%%%%%%%%%%%%%%%%%
\section{Skill Correlations}\label{sec_empiric}
%%%%%%%%%%%%%%%%%%%%%%%%%%%%%%%%%%%%%%%%
%%%%%%%%%%%%%%%%%%%%%%%%%%%%%%%%%%%%%%%%

The previous section highlighted the importance of understanding skill correlations in the workforce of interest. Here, we review the empirical studies that have estimated skill correlations (Section \ref{subsec_empiric_litreview}), present new evidence on the distribution of skill correlations across sectors (Section \ref{subsec_empiric_nlsy}), and show how positive assortative matching in an O-ring model of production (i.e., \citealt{kremer1993ring}) can generate negative skill correlations within firms even when the population exhibits a positive skill correlation (Section \ref{subsec_empiric_oring}).

%%%%%%%%%%%%%%%%%%%%%%%%%%%%%%%%%%%%%%%
\subsection{Existing Estimates of Skill Correlations}\label{subsec_empiric_litreview}
%%%%%%%%%%%%%%%%%%%%%%%%%%%%%%%%%%%%%%%
At least since \cite{willis1979education}, it has been appreciated that a single index of workers' skill will abstract away from important variation in the labor supply. However, empirical estimates of workers' skills on specific types of tasks remains a relatively challenging pursuit. Approaches to date typically involve accessing confidential data, intense measurement, or the estimation of structural models of labor supply.

Most studies examining multi-dimensional skills have focused on relatively aggregate categories: cognitive skills, social skills, and manual skills. While these categories may be much broader than the connotation of the ``tasks'' in our model, the logic is the same. Furthermore, these studies generally focus on large populations of workers (e.g., national statistics).

Broadly speaking, among these three categories of skills, cognitive and social skills show the most positive correlation. Estimates of the correlation in these skills spans from approximately 0.1 up to 0.7 (\citealt{mayer2008human,baker2014eyes,deming2017growing,guvenen2020multidimensional,lise2020multidimensional,girsberger2022interpersonal,barany2025two}). Cognitive and manual skills exhibit negative to weakly positive correlations, roughly in the range of --0.4 to 0.1 (\citealt{lindenlaub2017sorting,lise2020multidimensional,barany2025two}). Social and manual skills consistently show the most negative correlations of these categories, on the scale of -0.4 to -0.6 (\citealt{lise2020multidimensional,girsberger2022interpersonal,barany2025two}). Within the cateogry of cognitive skills, data persistently reveal a strong positive correlation, for example, of approximately 0.7 across mathematical skills and literacy (or verbal) skills (\citealt{hampf2017skills,guvenen2020multidimensional,woessmann2024skills}).

%%%%%%%%%%%%%%%%%%%%%%%%%%%%%%%%%%%%%%%%
\subsection{SAT Score Correlations across the Economy}\label{subsec_empiric_nlsy}
%%%%%%%%%%%%%%%%%%%%%%%%%%%%%%%%%%%%%%%%
As noted above, prior work has generally found mathematical and verbal skills to be positively correlated per standardized tests. A common source of data used is the National Longitudinal Survey of Youth (NLSY), which reports participants' SAT scores for verbal and math components separately (\citealt{bls2024nlsy97}). Here, we also make use of this data, but report the math-verbal skill correlation separately for alternative industries and occupations. We use the most commonly listed industry and occupation codes for all employed individuals. The public data yield 1,618 observations with both math and verbal SAT scores as well as non-missing industry and occupation codes.\footnote{For simplicity and to preserve sample size, we aggregate the industries and occupations into levels slightly broader than listed in the data. SATs are graded out of 800 for each of the two components. For confidentiality, the scores are aggregated into 100 point bins in the public data, which are what we use in this analysis. All correlations are reported per the inclusion of the NLSY97 sample weights, although this makes little practical difference.} 

Overall, math and verbal SAT scores in our sample exhibit a correlation of 0.61, which is squarely in-line with the estimates summarized in the previous subsection. But our primary goal is to shed some initial light on portions of the economy where skills (at least per this metric) may be more or less correlated. 

\begin{figure}[h!] \centering
\caption{Math and Verbal SAT Score Correlations across the Economy}\label{fig_rho_indocc}
\begin{subfigure}[t]{0.32\textwidth} \caption{Across\\Industries}\label{fig_rho_ind_major}
\includegraphics[width=\textwidth, trim=0mm 10mm 0mm 0mm, clip]{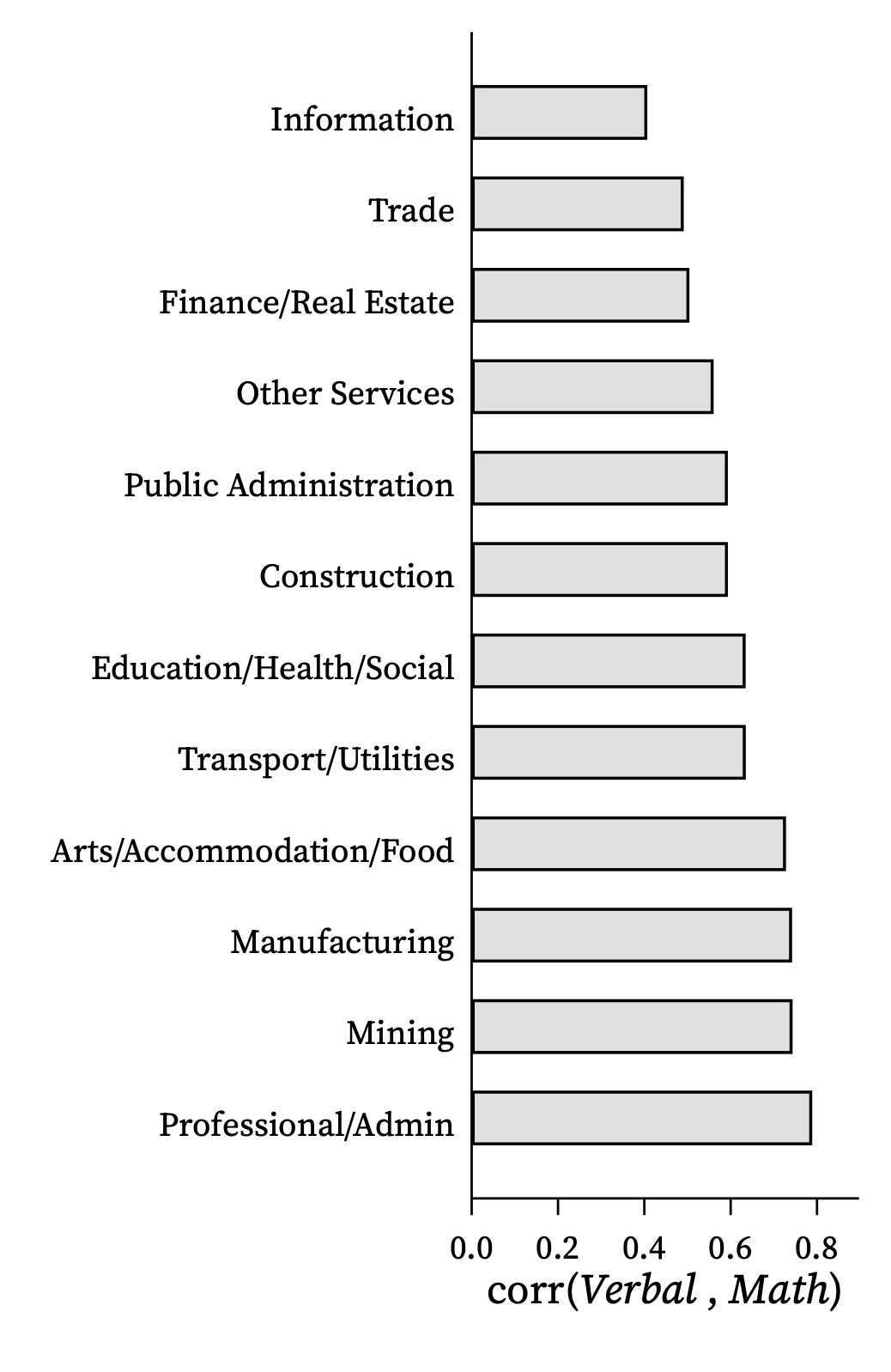}
\end{subfigure} \hfill \begin{subfigure}[t]{0.32\textwidth} \caption{Across \\Occupations}\label{fig_rho_occ_major}
\includegraphics[width=\textwidth, trim=0mm 10mm 0mm 0mm, clip]{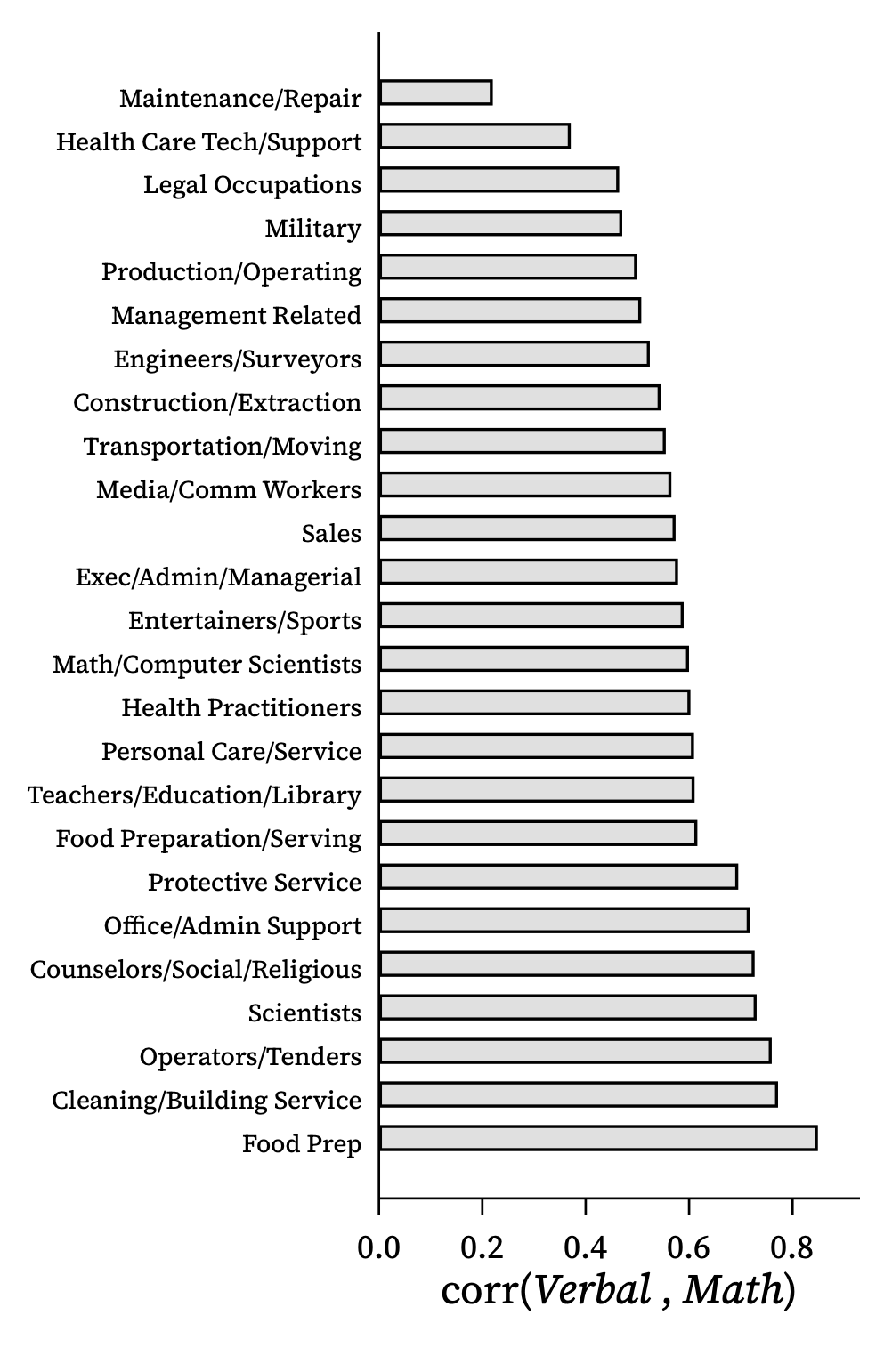}
\end{subfigure} \begin{subfigure}[t]{0.32\textwidth} \caption{Across\\Industry$\times$Occupations}\label{fig_rho_indocc_major}
\includegraphics[width=\textwidth, trim=6mm 20mm 0mm 0mm, clip]{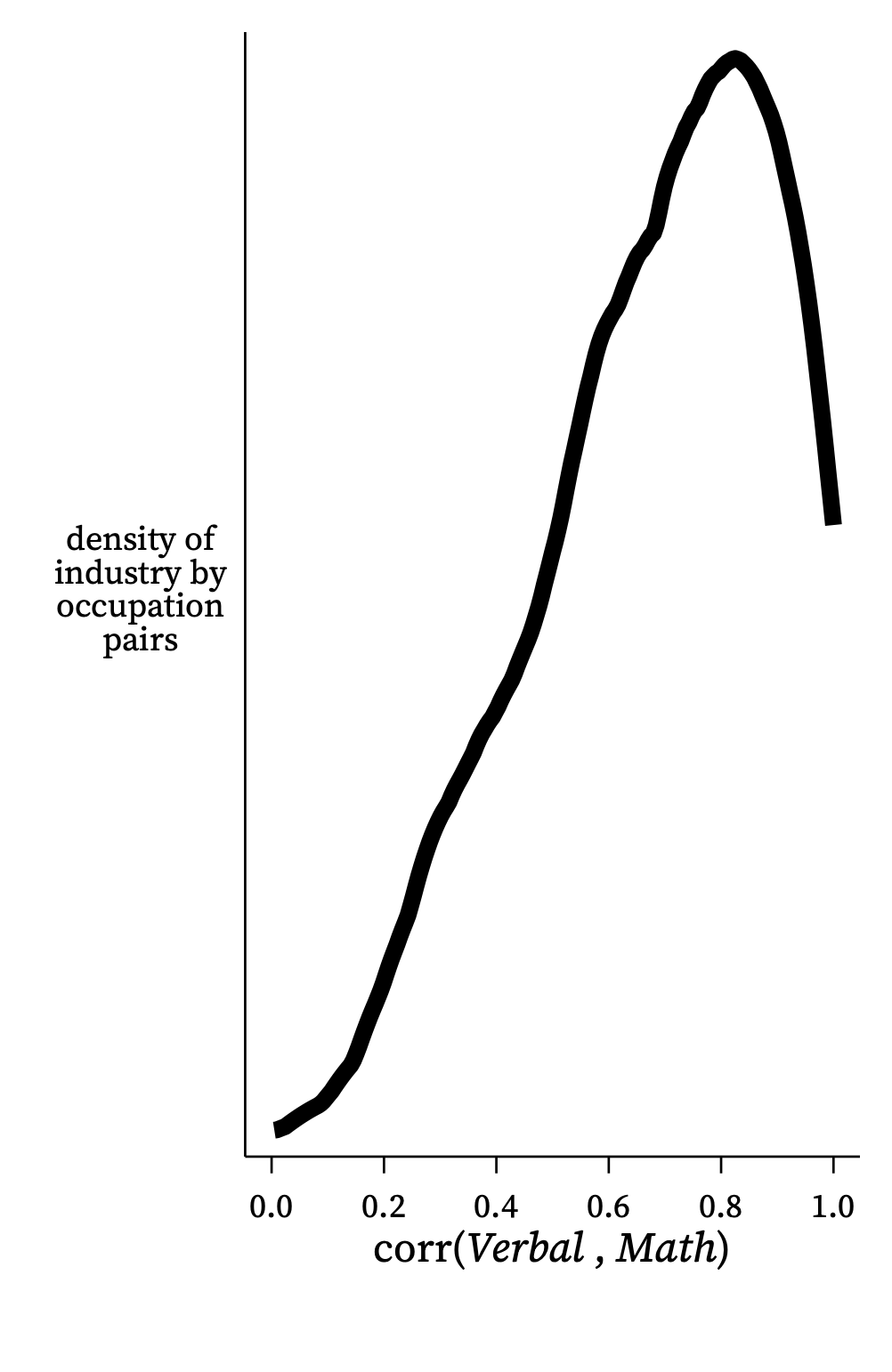}
\end{subfigure}
\note{\emph{Note:} Plots the correlation of individuals' SAT scores on the verbal and math components from the NLSY97. Panels (A--B) plot average correlations at the industry- (Panel A) or occupation-level (Panel B). Panel (C) plots the distribution of correlations averaged at the industry-occupation-level.}
\end{figure}

Figure \ref{fig_rho_indocc} highlights that the strength of the math-verbal score correlation varies significantly across the economy. 

Workers in some place exhibit a tight link between these skills, with estimated correlations approaching 0.8 (e.g., professional and administrative services, food preparation, cleaning and building services). While in other places the association is much more modest, on the order of 0.3 to 0.5 (e.g., information sector, legal services). At the granular level of industry–occupation pairs, we find substantial dispersion in math–verbal correlations (see Panel C).

Notably, we do not observe any industry and/or occupation where the correlation is estimated to be negative. Still, our model suggests that the the heterogeneity in skill correlations across the economy would also imply heterogeneous effects of automation on inequality. That is, a single automation technology deployed in these different sectors could yield significantly different changes in inequality.

More speculatively, the uniformly positive correlations help reconcile the fact that, as of this paper, the majority of automation experiments are documenting inequality defines. Presently, many of the technologies in question are both: ($i$) not yet ``super-human'' in that they are better than the worst workers, but not better than the best workers; and ($ii$) focused on cognitive tasks, in particular, connected to software. Combined with positive (cognitive) skill correlations, our model predicts an inequality reduction in this scenario. 

However, the positive math-verbal skill correlations we document here reflect sorting into broad economic sectors and may mask negative correlations at finer levels of analysis within firms or teams where complementary specialization drives task allocation.

%%%%%%%%%%%%%%%%%%%%%%%%%%%%%%%%%%%%%%%%
\subsection{An O-Ring Model with Two Tasks}\label{subsec_empiric_oring}
%%%%%%%%%%%%%%%%%%%%%%%%%%%%%%%%%%%%%%%%
The evidence in the prior section largely describes skill correlations in the aggregate (e.g., country-level). However, the focus of many automation experiments is there effect on inequality \emph{within} firms. Thus, we want to predict how the sorting of workers into firms may lead the within-firm skill correlation to diverge from the population correlation. To do so, we extend \citeauthor{kremer1993ring}'s (\citeyear{kremer1993ring}) O-ring model of production to the case where workers perform two tasks. This is related to automation experiments as it will emphasize how skill correlations (which drive inequality effects) can depend on the level of the workforce involved in the experiment (e.g., a business unit, a department, a firm, a sector, etc.). The model is similar to our focal model described in Section \ref{sec_model_setup}, although we make some additional assumptions in order to simply incorporate worker sorting and firm-level production.

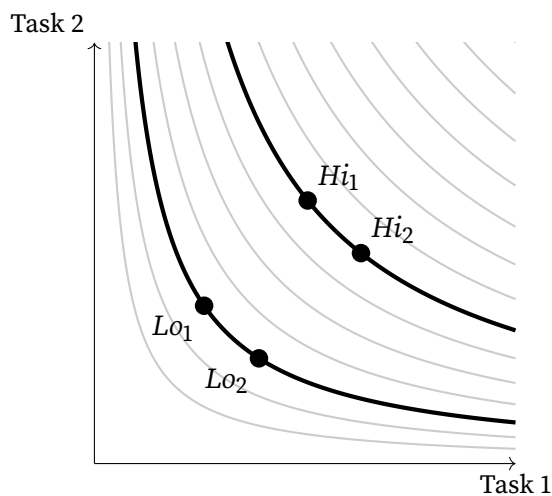
\begin{figure}[h!] \caption{Skill Correlations and Assortative Matching}\label{fig_isoqoring}
\begin{tikzpicture}[scale=0.7]
    % Axes
    \draw[->] (0,0) -- (8,0) node[below] {\small Task 1};
    \draw[->] (0,0) -- (0,8) node[above left] {\small Task 2};

    % Keep curves inside a clean 
    \begin{scope}
    \clip (0,0) rectangle (8,8);

    % Isoquants: X2 = (Ybar^2)/X1
    \foreach \Y in {1.5,2,2.5,3,3.5,4,4.5,5,5.5,6,6.5,7,7.5,8,8.5}{
    \draw[thick,black!20!white,domain=0.2:8,samples=300]
        plot (\x,{(\Y*\Y)/\x});
    }

    % Isoquants: X2 = (Ybar^2)/X1
    \foreach \Y in {2.5,4.5}{
    \draw[ultra thick,black,domain=0.2:8,samples=300]
        plot (\x,{(\Y*\Y)/\x});
    }

    \fill[black] ({(2.5^2)/3},{3}) circle (5pt) node[below left,scale=1.0]{$Lo_1$};
    \fill[black] ({(2.5^2)/2},{2}) circle (5pt) node[below left,scale=1.0]{$Lo_2$};
      
    \fill[black] ({(4.5^2)/5},{5}) circle (5pt) node[above right,scale=1.0]{$Hi_1$};
    \fill[black] ({(4.5^2)/4},{4}) circle (5pt) node[above right,scale=1.0]{$Hi_2$};
    
    \end{scope}
\end{tikzpicture}
\note{\emph{Note:} Plots isoquants of production possibilities, highlighting the isoquant for workers with higher ($Hi$) or lower output ($Lo$). Workers are chosen to illustrate a scenario where skills are positively correlated in the population (including all four workers). Since there are only two types of workers (and each type exists on the same isoquant), positive assortative matching in the context of two-worker firms will generate a negative within-firm skill correlation.}
\end{figure}

The intuition of this model can be seen clearly through Figure \ref{fig_isoqoring}. We've re-drawn the isoquants from Figure \ref{fig_isoq}, but now highlight the case where there are two workers of each type (high-output $Hi$ and low-output $Lo$). If firms consist of two workers and there is positive assortative matching (à la \citealt{kremer1993ring}), then workers of the same type will join. Then, because workers of the same type lie on the same isoquant in this stylized example, there will be a (perfectly) negative skill correlation within each of the two firms.\footnote{Conversely, negative assortative matching yields positive within-firm skill correlations in Figure \ref{fig_isoqoring}.} Next, we formalize this logic.

Consider an economy with a continuum of workers and firms. Each firm employs $n$ workers, and each worker performs two tasks. Worker $i$ has task-specific skill levels $L_{i1}, L_{i2} \in (0,1]$. Worker $i$'s production function is again the simple Cobb-Douglas form: $Y_i = \sqrt{L_{i1} \times L_{i2}}$.\footnote{We focus on Cobb-Douglas production in this analysis because the general CES form leads to non-linear constraints that do not have straightforward analytical solutions.} Following \cite{kremer1993ring}, firm-level output ($Z$) is the simple product of worker-level output:
\begin{equation}
Z = \prod_{i=1}^n Y_i \;,
\end{equation}
where we have abstracted away from capital used in production for simplicity, and because we are not interested in metrics related to, for example, labor shares.

Skills are distributed jointly log-normal in the population. Since our focus is on the correlation in skills, we'll focus on the simple case where the means ($\mu$) and standard deviations ($\sigma$) of both skills are equal:
\begin{equation}
\left(\;\begin{matrix} \ln L_{i1} \\ \ln L_{i2} \end{matrix} \;\right)
\sim \mathcal{N}\left(\;\begin{matrix} \mu \\ \mu \end{matrix}\;\;,\;\; 
\begin{matrix} \sigma^2 & corr_{\text{pop}}\sigma^2 \\ 
corr_{\text{pop}}\sigma^2 & \sigma^2 \end{matrix}\;\right) \;,
\end{equation}
where $corr_{\text{pop}} \in (-1,1)$ is the population correlation of logged skills.

The multiplicative firm-level structure ensures supermodularity in worker-level output: $\partial^2 Z/\partial Y_i \partial Y_j > 0$ for $i \neq j$. As in \cite{kremer1993ring}, equilibrium features positive assortative matching on the scalar index $Y_i$. 

Now, we can derive the relationship between the population-level skill correlation ($corr_{\text{pop}}$) and the within-firm skill correlation, which we'll denote with $corr_{\text{firm}}$, after workers sort into firms. In Appendix \ref{sec_app_derive} we solve for the equilibrium, within-firm correlation in skills and obtain:
\begin{equation}
    corr_{\text{firm}} = -\exp\left(-\frac{\sigma^2(1-corr_{\text{pop}})}{2}\right) < 0\;.
\end{equation}

In this example, the within-firm skill correlation is always negative regardless of the population-level correlation. The sharpness of this result stems from the assumptions of (multiplicative) complementarity among all workers within the firm and perfect positive assortative matching. As we highlighted above with Figure \ref{fig_isoqoring}, this leads to firms where all workers produce on the same isoquant, which, by definition, yields a negative skill correlation.\footnote{In a sense, this is an example of Simpson's paradox arising due to matching.} 

Certainly, the presence of substitutable workers within firms or imperfect positive assortative matching across firms would attenuate this effect.\footnote{For example, with random matching, the within-firm and population-level skill correlations will become equal.} Thus, in practice, within-firm skill correlations need not always be negative. Still, this highlights the importance of understanding how workers sort into different levels of aggregation when predicting skill correlations. This, in turn, implies that the scope of an automation experiment can have an important effect on the skill correlation of the workers in the experiment. Further exploration of worker sorting and firm-level production that accounts for these sorts of effects on skill correlations could prove useful.
%%%%%%%%%%%%%%%%%%%%%%%%%%%%%%%%%%%%%%%%
%%%%%%%%%%%%%%%%%%%%%%%%%%%%%%%%%%%%%%%%
\section{Discussion}\label{sec_discuss}
%%%%%%%%%%%%%%%%%%%%%%%%%%%%%%%%%%%%%%%%
%%%%%%%%%%%%%%%%%%%%%%%%%%%%%%%%%%%%%%%%

We have four main results. First, the inequality effects of automation depend on the interaction between skill correlations and the technology's capability relative to worker skills, rather than on the workforce's absolute skill level. Second, these effects are often non-monotonic: as the technology improves, inequality may first decrease and then increase, or vice versa. Third, the fastest path to equality (with respect to non-automatable tasks) is balanced technological progress across tasks. Fourth, positive assortative matching can generate negative, within-firm skill correlations even when the population-level correlation is positive.

Practically, this all implies that the same automation technology can reduce inequality in one experimental sample but increase it in another, even if both samples share identical production functions and skill distributions, so long as their underlying skill correlations differ. Moreover, because technology capabilities improve rapidly, an experiment showing an inequality reduction today may not imply an inequality reduction tomorrow.

For researchers planning future automation experiments, our framework suggests several priorities. First is the measurement of the correlation structure of participants' skills across tasks. Second is the technology's capability, not just in absolute terms, but relative to worker skill on each task.

Our simple model of automation experiments highlights a relatively unique feature of the workforce: the correlation of workers' skills across tasks. While many discussions of emerging automation technologies focuses on how workers with different \emph{levels} of skill are effected, we emphasize here that it is the \emph{correlation} in skills across tasks that is key for understanding how automation affects inequality.

The model's simplicity is both a strength and a limitation. By abstracting from general-equilibrium adjustments, we isolate the direct productivity channel that experiments measure. This modeling choice makes our predictions directly comparable to experimental results, but it leaves the model silent on longer-run inequality effects in general equilibrium. Incorporating skill correlations into macroeconomic models of automation is a natural next step to explore these broader consequences.

We have also abstracted from heterogeneity in workers' ability to use automation technologies effectively. If ``skill in using technology'' is itself a task and happens to be negatively correlated with other skills, our model's predictions would become less clear-cut. Recent experimental evidence that some workers perform worse with algorithmic assistance than without it (e.g., \citealt{dell2025knowledge,otis2025uneven}) indicates that this is an important extension to consider in future work.

Empirically, our investigation of SAT math and verbal scores of reveal substantial heterogeneity in cognitive skill correlations across industries and occupations; however, all correlations are positive. In domains where skills are positively correlated and the technology is not (yet) superhuman, our model predicts automation will disproportionately help lower-skill workers, compressing inequality. This is what most automation experiments focused on generative technologies have found to date. But as technologies improve toward and eventually beyond the highest human skill levels, our model predicts that their impact on inequality could reverse, with inequality beginning to rise.

More broadly, our framework suggests that the diversity of automation technologies will play a key role in the evolution of inequality. Technologies that excel at a narrow set of tasks may amplify inequality by primarily benefiting workers who specialize in those tasks or in complementary skills. In contrast, technologies with moderate capabilities across many tasks may compress inequality by helping workers overcome their weakest skills. This observation motivates further theoretical and empirical work on how the portfolio of available automation tools shapes the returns to various skill combinations.

Finally, our focus on task-level production and skill correlations connects to broader questions about the organization of work. When skills are negatively correlated across workers, firms have an incentive to exploit comparative advantage by assigning each worker to the tasks where their relative skill is highest. When skills are positively correlated, such specialization opportunities are limited, since the same workers tend to outperform others in all tasks. The interaction between skill correlations, automation, and the endogenous organization of production is a promising direction for future research.

%%%%%%%%%%%%%%%%%%%%%%%%%%%%%%%%%%%%%%%%
%%%%%%%%%% Bibliography
%%%%%%%%%%%%%%%%%%%%%%%%%%%%%%%%%%%%%%%%
\clearpage
\bibliography{\bib}

% %%%%%%%%%%%%%%%%%%%%%%%%%%%%%%%%%%%%%%%%
% %%%%%%%%%% Appendices
% %%%%%%%%%%%%%%%%%%%%%%%%%%%%%%%%%%%%%%%%
\newpage
\appendix

% %%%%%%%%%% APPENDIX A
\section{Derivation of Within-Firm Skill Correlations}\label{sec_app_derive}
\setcounter{table}{0}
\setcounter{figure}{0}
\setcounter{equation}{0} 
\renewcommand{\theequation}{A\arabic{equation}}
\renewcommand{\thetable}{A\arabic{table}}
\renewcommand{\thefigure}{A\arabic{figure}}
To summarize the O-ring model setup, each firm employs $n$ workers, and each worker $i$ performs two tasks using their task-specific endowments of skill: $L_{i1}, L_{i2} \in (0,1]$. Worker $i$'s production function is Cobb-Douglas and uses equal shares of both tasks. Following \cite{kremer1993ring}, firm-level output ($Z$) is the product of worker-level output: $Z = \prod_{i=1}^n Y_i$, which generates positive assortative matching of workers into firms. Skills are distributed such that:
\begin{equation*}
\left(\;\begin{matrix} \ln L_{i1} \\ \ln L_{i2} \end{matrix} \;\right)
\sim \mathcal{N}\left(\;\begin{matrix} \mu \\ \mu \end{matrix}\;\;,\;\; 
\begin{matrix} \sigma^2 & corr_{\text{pop}}\sigma^2 \\ 
corr_{\text{pop}}\sigma^2 & \sigma^2 \end{matrix}\;\right) \;,
\end{equation*}
where $corr_{\text{pop}} \in (-1,1)$ is the population correlation of logged skills.

With Cobb-Douglas production, $Y_i=(L_{i1} \times L_{i2})^{1/2}$. Within each firm $k$, all workers produce output $Y_k$, so $L_{i1} \times L_{i2} = Y_k^2$ for all workers $i$ in firm $k$. Taking logarithms of the constraint yields: $\ln L_{i1} + \ln L_{i2} = 2\ln Y_k$. All within-firm variation in logged skills comes from the difference in logged skill levels, which allows us to write:
\begin{equation}
\begin{aligned}
    \text{Var}(\ln L_{i1}-\ln L_{i2} \,|\, Y_k) & = \text{Var}(\ln L_{i1}-\ln L_{i2})\\
    & = 2\sigma^2(1-corr_{\text{pop}}) \;.
\end{aligned}
\end{equation}
Within each firm, we can express each logged skill levels as: $\ln L_{i1}=\ln Y_k+(\ln L_{i1}-\ln L_{i2})/2$ and $\ln L_{i2}=\ln Y_k-(\ln L_{i1}-\ln L_{i2})/2$. Therefore:
\begin{equation}
\begin{aligned}
    \text{Var}(\ln L_{i1} \,|\, Y_k) & = \text{Var}(\ln Y_k+(\ln L_{i1}-\ln L_{i2})/2 \,|\, Y_k) \\
    & = \text{Var}((\ln L_{i1}-\ln L_{i2})/2) \\
    & = \frac{\sigma^2(1-corr_{\text{pop}})}{2} \;,
\end{aligned}
\end{equation}
and likewise for $\text{Var}(\ln L_{i2} \,|\, Y_k)$. Now, transforming back to levels, we have $L_{i1}=Y_k \times \exp((\ln L_{i1}-\ln L_{i2})/2)$ and $L_{i2}=Y_k \times \exp(-(\ln L_{i1}-\ln L_{i2})/2)$, where $(\ln L_{i1}-\ln L_{i2})/2 \sim \mathcal{N}(0, \sigma^2(1-corr_{\text{pop}})/2)$. Therefore:
\begin{equation}
\begin{aligned}
    \text{Var}(L_{i1} \,|\, Y_k) & = Y_k^2 \times \exp\left(\frac{\sigma^2(1-corr_{\text{pop}})}{2}\right)\left(\exp\left(\frac{\sigma^2(1-corr_{\text{pop}})}{2}\right)-1\right) \;,
\end{aligned}
\end{equation}
and likewise for $\text{Var}(L_{i2} \,|\, Y_k)$.
The covariance is then:
\begin{equation}
\begin{aligned}
    \text{Cov}(L_{i1}, L_{i2} \,|\, Y_k) & = Y_k^2 \times \left(1-\exp\left(\frac{\sigma^2(1-corr_{\text{pop}})}{2}\right)\right) \;,
\end{aligned}
\end{equation}
and, finally, the within-firm correlation is:
\begin{equation}
\begin{aligned}
    corr_{\text{firm}}  & = \frac{\text{Cov}(L_{i1}, L_{i2} \,|\, Y_k)}{\sqrt{\text{Var}(L_{i1} \,|\, Y_k)\times\text{Var}(L_{i2} \,|\, Y_k)}} \\
    & = \frac{1-\exp\left(\frac{\sigma^2(1-corr_{\text{pop}})}{2}\right)}{\exp\left(\frac{\sigma^2(1-corr_{\text{pop}})}{2}\right)\left(\exp\left(\frac{\sigma^2(1-corr_{\text{pop}})}{2}\right)-1\right)} \\
    & = -\exp\left(-\frac{\sigma^2(1-corr_{\text{pop}})}{2}\right) \;.
\end{aligned}
\end{equation}

\end{document}